# Checkerboard-type patterns of charge stripes in two-gap superconductor $ZrB_{12}$

N. B. Bolotina[1], O. N. Khrykina[1], A. N. Azarevich[2], N. Yu. Shitsevalova[3],

V. B. Filipov[3], S. Yu. Gavrilkin[4], K. V. Mitsen[4], N. E. Sluchanko[2]

[1] *Shubnikov Institute of Crystallography, Federal Scientific Research Centre 'Crystallography and Photonics' of Russian Academy of Sciences, 59 Leninskiy prospect, Moscow, 119333, Russia*

[2] *Prokhorov General Physics Institute of the Russian Academy of Sciences, Vavilov str. 38, Moscow 119991, Russia*

[3] *Frantsevich Institute for Problems of Materials Science, National Academy of Sciences of Ukraine, 03680 Kyiv, Ukraine*

[4] *Lebedev Physical Institute of the Russian Academy of Sciences, 53 Leninskiy prospect, 119991 Moscow, Russia*

**Abstract.**

Inhomogeneous superconductivity in the high quality single crystals of $ZrB_{12}$ ($T_c \approx 6$ K) has been studied using the heat capacity and x-ray diffraction (XRD) data. Evidence of two-band superconductivity with two branches of upper critical field $\mathbf{H}_{c2}(T_c)$ is obtained in a magnetic field applied along the [110] axis of the crystal. On the contrary, at $\mathbf{H} \parallel [100]$, the only dependence $\mathbf{H}_{c2}(T_c)$ is observed. This finding is supplemented with the checkerboard-type patterns of the charge stripes in $ZrB_{12}$ deduced from the detailed analysis of XRD data. These patterns are compared to the structure of the charge stripes in the weakly bound superconductor $LuB_{12}$, whose $T_c$ is 15 times lower than that of $ZrB_{12}$. Probable nature of the two-gap superconductivity in $ZrB_{12}$ with strongly enhanced characteristics is discussed.

**I. Introduction.**

In the study of high-temperature superconductivity (HTSC) in conventional ($MgB_2$[1]) and unconventional (cuprates, Fe-based pnictides and chalcogenides [2–6]) superconductors, many unusual phenomena were discovered, including charge and spin stripes [2–4], electron nematic effect [2], multiband superconductivity of various types (see [5-7] for review), etc. It is assumed that at least some of these features are closely related to the mechanisms of the superconductivity enhancement [2–7]; therefore, the elucidation of the nature of these anomalies and their relationship with HTSC is extremely important. It is believed at present that an interplay of simultaneously active charge, spin, lattice, and orbital interactions plays a key role in the formation of a rich variety of phases in the phase diagrams of these superconductors and provides the essential ingredients of HTSC [2-8].

Recently the question about the microscopic mechanism, which causes the material to form the charge stripes through the interplay of charge and lattice degrees of freedom, has been successfully clarified for the weakly coupled superconductor $LuB_{12}$ ($T_c \approx 0.42$ K). According to the authors of [9, 10], cooperative Jahn-Teller (JT) instability (ferrodistortive effect) of rigid boron framework is the factor responsible for the periodic changes in the 5$d$-2$p$ hybridization of the conduction band states, resulting in formation of *the dynamic charge stripes* in the <110> directions of the *fcc* crystal structure of $LuB_{12}$. When studying solid solutions $Lu_xZr_{1-x}B_{12}$, it was shown [11] that the charge fluctuations at the Lu sites caused an effect of pair breaking, leading to a 15-fold decrease in $T_c$ with $x$ increase. $ZrB_{12}$ shows itself as a *two-gap strongly coupled superconductor* with $T_c \approx 6$ K (see [11] for review), but the reason for the $T_c$ enhancement and variation of the electron-phonon interaction $\lambda_{e-ph}$ in the range 0.4 – 1 is not yet clear. Taking into account that a large *pseudo-gap* ($\Delta_{ps-gap} \approx 7.3$ meV) has been detected employing high resolution photoemission spectroscopy in $ZrB_{12}$ above $T_c$, and *proximity to the quantum fluctuation regime* has been predicted from the *ab initio* calculations of the band structure [12], it is promising to study both the fine details of the crystal and electronic structure, and the anisotropy of the superconducting state in the two-gap $ZrB_{12}$ superconductor in order to find patterns that may be important for HTSC.

Here the crystal structure of $ZrB_{12}$ is studied at room and low temperatures. Small static Jahn-Teller distortions of the *fcc* lattice are observed. The detected charge stripes form *two grids from rhomboid cells (checkerboard patterns)* constructed from (i) hybridized 4$d$-2$p$- and (ii) only 2$p$-conduction band states. The characteristics of the two-gap superconductivity are testified by measurements of the heat capacity. We conclude in favor of the magnetic field-induced anisotropy, which is due to the orientation of the two-band filamentary structure of the electron density.

**II. Experimental results.** The high quality single crystals of $ZrB_{12}$ were cut from the same rods as in [11]. The heat capacity measurements were carried out on PPMS-9 (Quantum Design). The x-ray data sets (Mo$K_\alpha$ radiation, $\lambda = 0.7093$ Å) were obtained at 97 K and at room temperature 293 K using an Xcalibur diffractometer (Oxford Diffraction). To cool the sample, a Cobra Plus cryosystem (Oxford Cryosystems) was applied with an open stream of cold nitrogen directed at the sample. Crystal structure of $ZrB_{12}$ was refined in the $Fm\bar{3}m$ group using Jana2006 program [13]. Table 1 contains the *fcc* lattice parameters, equivalent atomic displacement parameters, and selected interatomic distances for $ZrB_{12}$ at two temperatures. The unit cell of the *fcc* lattice contains two symmetrically independent atoms: Zr in the fixed position 4$a$ (0, 0, 0) and B in the special position 48$i$ (1/2, $y$, $y$), $y \approx 1/6$. Each Zr atom is surrounded by 24 boron atoms and 12 Zr atoms, $d$(Zr–Zr) = $a_{cub}/\sqrt{2} \approx 5.23$ Å. Other details of the data collection and structure refinement are summarized in Table S1 [14].

**Table 1.** The *fcc* lattice parameter ($a_{cub}$); $y$-coordinate of B; equivalent atomic displacements ($u_B$ and $u_R$); distances $d$(Zr – B), $d$(B – B)$_{intra}$ between the neighboring boron atoms in one cuboctahedron, $d$(B – B)$_{inter}$ between nearest boron atoms in neighboring cuboctahedra in ZrB$_{12}$ at room and low temperatures

| $T$, K | $a_{cub}$, Å | $y_B$ | $u_B$, Å$^2$ | $u_{Zr}$, Å$^2$ | $d$(Zr-B), Å | $d$(B-B)$_{intra}$ | $d$(B-B)$_{inter}$ |
|---|---|---|---|---|---|---|---|
| 293 | 7.40203(2) | 0.16977(2) | 0.00455(2) | 0.00539(3) | 2.7485(2) | 1.7772(2) | 1.6797(3) |
| 97 | 7.3985(1) | 0.16966(5) | 0.00324(4) | 0.00256(6) | 2.7475(4) | 1.7752(4) | 1.6811(5) |

**II.1. Crystal and electron structure.**

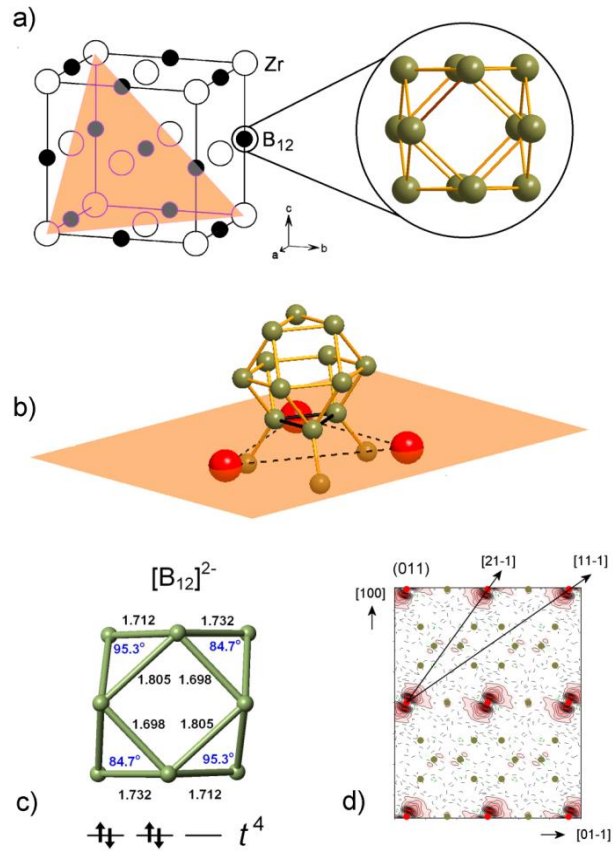

**Figure 1**. (**a**) Schematic representation of the NaCl-type unit cell of ZrB$_{12}$. Open circles are Zr atoms, dark circles are B$_{12}$ cuboctahedra, one of which is shown on the right. The (111) plane is highlighted. (**b**) Three Zr atoms (large red balls) and six boron atoms (small balls, orange below and khaki above the (111) plane) form three Zr-B-B-Zr bridges. (**c**) Theoretically calculated JT distortions of a negatively charged [B$_{12}$]$^{-2}$ cluster. The cuboctahedron B$_{12}$ is shown in projection onto one of its square faces. (**d**) Difference Fourier synthesis of electron density (ED) in the (011) plane at $T$ = 97 K. Red circles are Zr sites in the plane, khaki-colored circles are boron sites at distances less than 1 Å from the plane. Positive ED residues are shown in shades of red. Contour interval is 0.5 e/Å$^3$.

The crystal structure of $ZrB_{12}$, typical for most dodecaborides, is schematically presented in fig. 1a. As shown in [15], charge transfer across the bridge between two Zr atoms is possible through a pair of boron atoms owing to significant overlapping at Fermi level ($E_F$) between the B – B π-bond and 4$d$ orbital of Zr. Charge transfer along the Zr-B-B-Zr bridges is possible in six directions <110>, which are equivalent in a cubic structure. The plane (111) in fig. 1b contains three such directions indicated by dashed lines. Boron atoms forming three B-B pairs go out of the plane by ±0.8 Å. Three atoms, one from each pair, form a triangular face of the $B_{12}$ cuboctahedron, while the other three belong to three different cuboctahedra.

The geometry of the negatively charged clusters $[B_{12}]^{n-}$ was earlier [10, 16] optimized within the framework of the density functional theory (DFT), which made it possible to conclude in favor of their JT distortions (see Section 1 in [14]). The result of DFT calculations for the cluster $[B_{12}]^{2-}$, which is the element of $ZrB_{12}$ crystal structure, is shown in fig. 1c. Parallel alignment of local distortions of $B_{12}$ cubooctahedra (known as the *ferrodistortive* case) can lead to static and dynamic JT deformations of the *fcc* lattice as a whole. Indeed, when the lattice parameters of $ZrB_{12}$ were refined without imposing symmetry bonds, the very small but well-detected static JT distortions of the cubic lattice were found both at 293 K and 97 K (less than 0.01 Å in length and 0.02° in angles). With such small violations, it makes no sense to lower the symmetry of the structural model; all our attempts to do this were ineffective. At the same time the *static lattice distortions* are a fingerprint of the *dynamic JT instability* [10,17]. This leads to an asymmetric redistribution of electrons on metal-boron and boron-boron bonds in the interstices of the crystal lattice. Their visualization on the difference Fourier maps of electron density (ED) is provided by a simple technique (see Section 2 in [14]), which was applied previously for other JT-active dodecaborides $LuB_{12}$ [9, 10] and $(Tm,Yb)B_{12}$ [10, 18]. Positive residuals of electron density near Zr sites in $ZrB_{12}$ form a flat, highly anisotropic butterfly-shaped figure with wings parallel to the (011) plane (fig. 1d). If interatomic distances are calculated from the lattice parameters distorted by the JT effect, cuboctahedra $B_{12}$ are slightly elongated along [11-1] and compressed in the perpendicular plane (11-1), as was theoretically predicted (fig. 1c) for the ferrodistortive JT state [10, 16]. This effect persists at room temperature, but is partially suppressed by the thermal motion of the atoms.

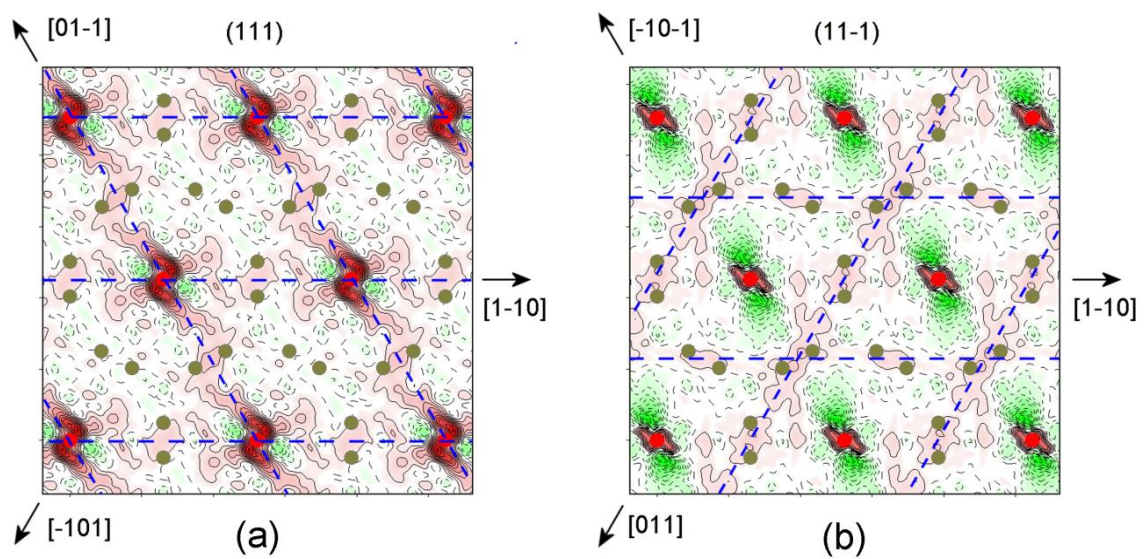

**Figure 2**. Distribution of difference electron density in a 12 × 12 Å fragment of (a) (111) and (b) (11-1) plane of the ZrB$_{12}$ crystal lattice at $T$ = 97 K. Red circles are Zr sites in the plane, khaki-colored circles are boron sites at distances less than 1 Å from the plane. Positive and negative ED residues are shown in shades of red and green, respectively. Contour intervals are 0.2 e/Å$^3$. The dotted blue lines are guides for eye demonstrating the stripe configurations in the planes.

The advantageous charge transfer along [01-1] across the Zr-B-B-Zr bridge is evident in the plane (011) from fig. 1d. In the (111) plane (fig. 2a), one more direction is distinguished in addition to [01-1], namely [1-10]. As follows from fig. 2b, charge transfer in the (11-1) plane along Zr-B-B-Zr bridges is less probable, but an excess of electrons is collected in the boron lattice, mainly along [011] and [1-10] directions. These results allow us to (i) conclude that these (111) and (11-1) planes become nonequivalent in the JT-active *fcc* lattice and (ii) make an assumption about the ways of charge transfer in ZrB$_{12}$. The conclusion by Ma *et al* [15] on the formation of Zr-B-B-Zr bridges along the <110> directions owing to overlapping at Fermi level between the B-B π-bond and 4*d* orbital of Zr should be supplemented with the assumption of additional charge transfer over the boron states bypassing 4*d*-orbitals of zirconium. Fine electron structure of LuB$_{12}$, previously studied in a similar way [9], demonstrated the only linear charge stripes running through the bridges Lu-B-B-Lu in this superconductor with $T_c \approx 0.42$ K.

**II.2. Two-gap superconductivity in ZrB$_{12}$.** The temperature dependencies of the heat capacity $C(T, H_0)$ of ZrB$_{12}$ single crystals recorded in the temperature range 0.4 – 7 K and in magnetic fields up to 3 kOe applied along [100] and [110] directions are shown in figs. 3a and 3b, correspondingly. It is certainly discernible in fig.3b that there are two adjacent humps (the main and smaller one) on the $C(T, H_0)$ curves due to the superconducting transitions in the field orientation **H** ∥ [110], but only one step-like singularity is observed for **H** ∥ [100] (fig.3a). Figs. 3c and 3d highlight the temperature dependences of thermodynamic $H_{cm}(T)$ and upper $H_{c2}(T)$, $H_{c2}'(T)$ critical fields (see Section 3 in [14] for more details) and the zero-field normalized heat capacity behavior $C/T = f(T/T_c)$ in the superconducting state, correspondingly. It is interesting to note from fig. 3c that in the range 4 – 6 K the smaller gap superconductivity for **H** ∥ [110] is type-I, but at $T^* \sim 4$ K the transition from type-I to type-II/1 is observed in the lower band. On the contrary the transition from type-II to type-II/1 is clearly discern at $T^* \approx 4.5$ K in the upper superconducting band and for **H** ∥ [110] (see also [11] for review).

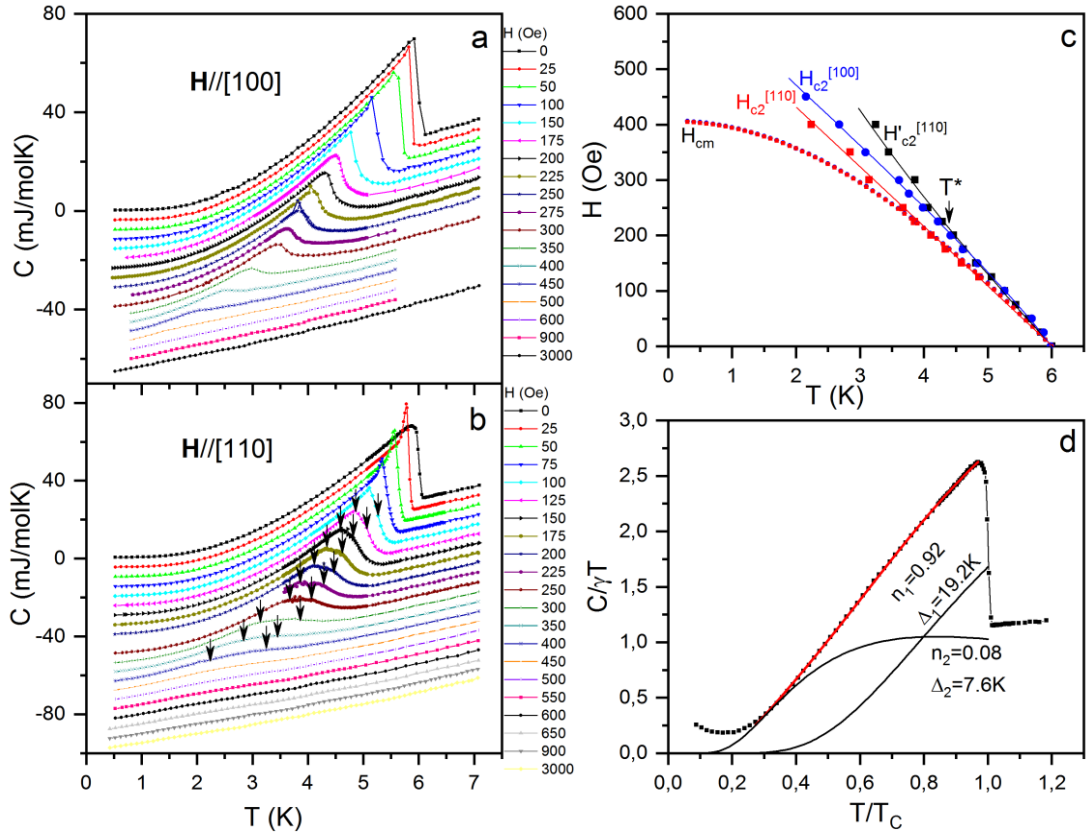

**Figure 3.** Temperature dependencies of the heat capacity $C(T, H_0)$ of $ZrB_{12}$ single crystals in magnetic fields $H_0 \leq 3$ kOe applied along **(a)** [100] and **(b)** [110] directions. Arrows mark two adjacent transitions. **(c)** Thermodynamic $H_{cm}(T)$ and upper $H_{c2}(T)$, $H_{c2}'(T)$ critical fields and **(d)** the zero-field normalized heat capacity $C/T = f(T/T_c)$ in the superconducting state (see text).

**III. Discussion.** The zero temperature critical fields $H_{cm}(0)$, $H_{c2}^{[110]}(0)$, $H_{c2}'^{[110]}(0)$ and $H_{c2}^{[100]}(0)$ deduced in the study (Section 3 in [14]) allow estimating Ginzburg-Landau-Maki (GLM) parameters $\kappa_1(0) \approx 1.05$, $\kappa_2(0) \approx 0.97$ for superconductivity in two different conduction bands. It is also worth noting a moderate difference in the coherence length $\xi_1(0) \approx 739$ Å, $\xi_2(0) \approx 766$ Å of the lower and upper bands. Using two-band $\alpha$-model fit (Section 4 in [14]) the gap energies $\Delta_1(0)/k_B \approx 19.2$ K and $\Delta_2(0)/k_B \approx 7.6$ K were calculated in combination with the relative weight $n_2(x) \approx 0.08$ of the small-gap component. The ratios $2\Delta_1(0)/k_B T_c \approx 6$ and $2\Delta_2(0)/k_B T_c \approx 2.5$ obtained here argue in favor of strong coupling and weak coupling superconductivity in the upper and lower bands, correspondingly.

The band structure calculations in the rigid *fcc* lattice of $ZrB_{12}$ [12, 19, 20] predict only hybridized $4d$-$2p$ conduction band states near $E_F$. On the contrary, static and dynamic JT distortions observed in $ZrB_{12}$ give rise to another conduction band constructed from $2p$ boron orbitals. It can be assumed that two-band superconductivity is characteristic of two-band metallic $ZrB_{12}$. Its conducting channels form two filamentary structures in the crystal, each in form of *a grid with rhomboid cells (a checkerboard pattern)*. One of them is generated by hybridized $4d$-$2p$ orbitals and other one – by $2p$ conduction band states. Note that the arrangement of the two types of plane *grids of stripes* could be considered as a factor responsible for the field induced anisotropy of superconductivity detected in the heat capacity study (fig.3c). Then, when comparing the conduction band of dodecaborides $LuB_{12}$ and $ZrB_{12}$ with a similar *fcc* structure and similar phonon spectra, one should take into account the difference in the valences of Zr (4+) and Lu (3+). Indeed, the only $5d$ electron of the $Lu^{3+}$ ion participates in the formation of $5d$-$2p$ hybridized conduction states, with the charge stripes located along the <110> directions in the *fcc* lattice [9, 10]. In contrast, each $Zr^{4+}$ ion in $ZrB_{12}$ gives two $4d$ electrons that contribute to the conduction states, which leads to both an increase in $E_F$ by about 0.3-0.4 eV [20] and the appearance of two checkerboard patterns in the filamentary structure of the conduction channels (fig. 2). It should be noted that, according to the conclusions [2-4], the charge stripes should be considered as important components in HTSC. Moreover in [21] a new scenario was proposed based on the formation of pair density waves (PDW) and charge density waves (CDW), which underlie the emergence of a complex inhomogeneous superconducting state in the HTSC. This kind of PDW, which intertwines both CDW and superconducting orders, can also be used to explain inhomogeneous superconductivity with the checkerboard patterns of charge stripes observed in $ZrB_{12}$.

## IV. Conclusion.

Precise XRD measurements of high-quality $ZrB_{12}$ crystals and the data analysis undertaken here establish both the static Jahn-Teller distortions of the *fcc* lattice and appearance of two types of checkerboard patterns of the charge stripes in the inhomogeneous superconductor. Our results of studying the low-temperature heat capacity allow us to conclude in favor of two-gap superconductivity and a noticeable anisotropy of the upper critical field $H_{c2}(T_c)$, which also depends on the orientation of the magnetic field vector. We estimate the GLM parameters $\kappa_1(0) \approx 1.05$, $\kappa_2(0) \approx 0.97$, the coherence lengths $\xi_1(0) \approx 739$ Å, $\xi_2(0) \approx 766$ Å, the energy gaps $\Delta_1(0)/k_B \approx 19.2$ K and $\Delta_2(0)/k_B \approx 7.6$ K, and ratios $2\Delta_1(0)/k_B T_c \approx 6$ and $2\Delta_2(0)/k_B T_c \approx 2.5$, which are the characteristics of superconductivity in the upper and lower bands with weak and strong coupling, correspondingly. We propose that the spatially modulated superconducting order in $ZrB_{12}$ could be discussed in terms of the PDW scenario developed recently for HTSC [21].

**Acknowledgements.**

This work was supported by the Ministry of Science and Higher Education within the State assignment FSRC 'Crystallography and Photonics' RAS in part of the data collection and structure refinement. The structure-property relationship was analyzed with the support of the Russian Foundation for Basic Research (Grant No. 18-29-12005). The diffraction data were collected using the equipment of the Shared Research Centre of FSRC 'Crystallography and Photonics' RAS with the support of by the Russian Ministry of Education and Science (project RFMEFI62119X0035); heat capacity measurements were performed at Shared Facility Centre of Lebedev Physical Institute. The authors are grateful to S. V. Demishev, V. V. Glushkov and S. J. Blundell for helpful discussions.

**Supplementary Information to the paper**

**Checkerboard-type patterns of charge stripes in two-gap superconductor ZrB$_{12}$**

N. B. Bolotina[1], O.N. Khrykina[1], A.N. Azarevich[2], N.Yu. Shitsevalova[3],

V.B. Filipov[3], S.Yu. Gavrilkin[4], K.V. Mitsen[4], N.E. Sluchanko[2]

[1] *Shubnikov Institute of Crystallography, Federal Scientific Research Centre 'Crystallography and Photonics' of Russian Academy of Sciences, 59 Leninskiy prospect, Moscow, 119333, Russia*

[2] *Prokhorov General Physics Institute of the Russian Academy of Sciences, Vavilov str. 38, Moscow 119991, Russia*

[3] *Frantsevich Institute for Problems of Materials Science, National Academy of Sciences of Ukraine, 03680 Kyiv, Ukraine*

[4] *Lebedev Physical Institute of RAS, 53 Leninskiy prospect, 119991 Moscow, Russia*

**Table S1**. Experimental details and refinement parameters for the ZrB$_{12}$ crystal structure.

| | 293 | 97 |
|---|---|---|
| $T$, K | 293 | 97 |
| Symmetry group | Fm-3m | |
| $a$, Å | 7.40203(2) | 7.3985(1) |
| $V$, Å$^3$ | 405.558(2) | 404.978(2) |
| $\mu$, мм$^{-1}$ | 2.514 | 2.518 |
| Diffractometer | Xcalibur EOS S2 CCD | |
| Radiation type; $\lambda$, Å | Mo$K_\alpha$; 0.70930 | |
| Scan type | $\omega$ | |
| $\theta_{max}$, degrees | 73.78 | 73.87 |
| No. of measured, independent and observed [$I > 3\sigma(I)$] reflections | 10479, 260, 260 | 10404, 260, 260 |
| Redundancy | 40.30 | 40.01 |
| $R_{int}$, %* | 5.83 | 7.34 |
| No of reflections (N) / No of parameters (P) | 260/8 | |
| $R(F) / wR(F)$, %** | 0.99/1.53 | 1.90/2.66 |
| $S$** | 0.96 | 1.00 |
| $\Delta\rho_{min}/\Delta\rho_{max}$, e/Å$^3$ | -0.49/0.67 | -1.24/1.21 |

*The internal factor $R_{int}$ estimates the average spread of intensities in groups of equivalent reflections:

$$R_{int} = \frac{1}{N} * \sum_{i=1}^{N} \sum_j \frac{(I_j - \bar{I}_i)}{\bar{I}_i} \text{ with } \bar{I}_i = \frac{\Sigma_j(I_j)}{j}$$

where *i* runs over *N* independent reflections with intensities $I_i$, and *j* – over all equivalent reflections corresponding to the *i*-th independent reflection.

** The residual factor $R$, the weighted residual factor $wR$ and the goodness-of-fit $S$ characterize the quality of the structural model refinement by the least squares method:

$$R(F) = \frac{\sum ||F|_{obs} - |F|_{calc}|}{\sum |F|_{obs}}; \quad wR(F) = \sqrt{\frac{\sum [w(|F|_{obs} - |F|_{calc})]^2}{\sum (w|F|_{obs}^2)}}; \quad S = \sqrt{\frac{\sum [w(|F|_{obs} - |F|_{calc})^2]}{N - P}}$$

**Section 1. DFT calculations**

Previously, quantum chemical calculations and geometry optimizations for the neutral cluster $[B_{12}]^0$ and negatively charged $[B_{12}]^{n-}$ clusters ($n$ = 1-4) were performed to evaluate probable JT distortions of the boron lattice in LuB$_{12}$ (Ref1, Ref2). Calculations were carried in terms of density functional theory (DFT) using ORCA 3.0.3 quantum chemistry program package at the BP86/def2-SVP level of theory. The initial point corresponded to the regular B$_{12}$ cuboctahedron and the structure was allowed to relax freely to reach the local minimum in the potential energy surface. Calculated structures of $[B_{12}]^{n-}$($n$ = 0-4) clusters indicated that the JT-active clusters $[B_{12}]^{n-}$($n$ = 0-3) were slightly distorted cuboctahedra. As expected, the non-degenerate $[B_{12}]^{4-}$ cluster had the $O_h$ symmetry of a regular cuboctahedron (fig. S1).

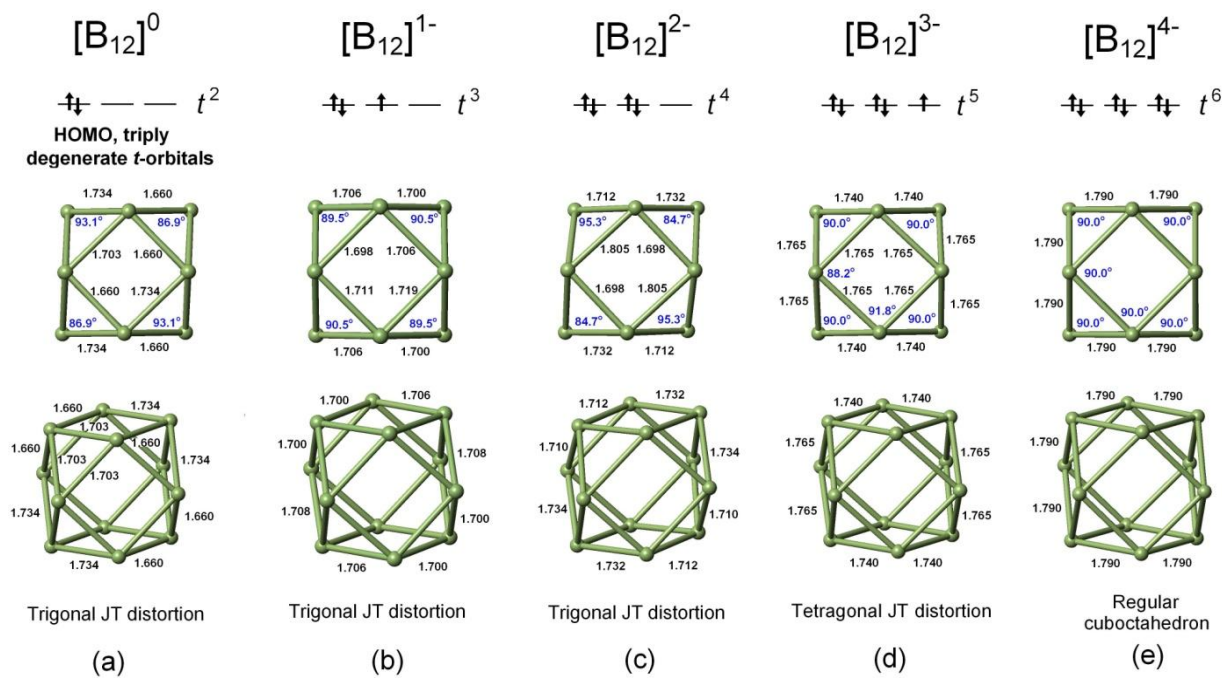

Fig. S1. Neutral and negatively charged isolated $[B_{12}]^{n-}$ clusters ($n = 0 - 4$) obtained from DFT calculations.

Ref1. N. Sluchanko, A. Bogach, N. Bolotina, V. Glushkov, S. Demishev, A. Dudka, V. Krasnorussky, O. Khrykina, K. Krasikov, V. Mironov, V. Filipov, and N. Shitsevalova, Phys. Rev. B **97**, 035150 (2018).

Ref2. N. B. Bolotina, A. P. Dudka, O. N. Khrykina, V. S. Mironov, Crystal structure of dodecaborides: complexity and simplicity, in *Rare-Earth Borides*, edited by D. S. Inosov (Jenny Stanford Publishing, Singapore, 2021), Chapter 3.

## Section 2. Difference Fourier synthesis of electron density

The visualization of charge stripes on the maps of the difference Fourier synthesis of electron density $\Delta\rho(\mathbf{r})$ is provided by a simple technique, although unusual for traditional structural analysis. $\Delta\rho(\mathbf{r})$ is calculated in any point of the unit cell using well-known formula:

$$\Delta\rho(\mathbf{r}) = \frac{1}{V} * \sum_{\mathbf{H}} [F_{obs}(\mathbf{H}) - F_{calc}(\mathbf{H})] \exp[-2\pi i (\mathbf{r} \cdot \mathbf{H})] \qquad (2.1)$$

Here $V$ is a unit-cell volume; $\mathbf{H} = \mathbf{H}(hkl) = \mathbf{S} - \mathbf{S}_0$ is a Bragg vector («reflection») where $\mathbf{S}_0$ and $\mathbf{S}$ are direct and diffracted x-ray beams, respectively. As one can see, this formula does not contain any information on the crystal symmetry. Structural factors $F_{calc}(\mathbf{H})$, which are necessary for the synthesis, are calculated from the structural parameters refined in the $Fm\bar{3}m$ group. Our departure from tradition concerns $F_{obs}(\mathbf{H}) = |F_{obs}(\mathbf{H})|\varphi_{calc}(\mathbf{H})$. Their phases $\varphi_{calc}(\mathbf{H})$ are taken as usual from $F_{calc}(\mathbf{H})$ with the same $\mathbf{H}$, but observed modulus $|F_{obs}(\mathbf{H})| \sim \sqrt{I}$ are not averaged in the Laue class $m\bar{3}m$, and the difference Fourier synthesis is not limited to a small unit-cell volume, independent in $Fm\bar{3}m$. Traditionally, synthesis is carried out in a symmetrically independent volume of a unit cell, and the result is extended to the entire cell using symmetry operators. In other words, the symmetry of the difference Fourier maps exactly corresponds to the space group supplied to the input of the synthesis procedure. Even if violations do occur, they will not appear on maps.

## Section 3. Thermodynamic and upper critical fields

The specific heat results obtained in the normal and superconducting states (Fig. 3 in the paper) were used to determine the thermodynamic critical field $H_{cm}(T)$ within the framework of standard relations

$$-1/2\ \mu_0 V H_{cm}^2(T) = \Delta F(T) = \Delta U(T) - T\Delta S(T) \qquad (3.1)$$

$$\Delta U(T) = \int [C_s(T') - C_n(T')]dT' \qquad (3.2)$$

$$\Delta S(T) = \int dT'[C_s(T') - C_n(T')]/T' \qquad (3.3),$$

where $F$ and $U$ denote the free and internal energies, $S$ the entropy, $V$ the molar volume, and the indices $n$ and $s$ correspond to characteristics of the normal and superconducting phases of $Lu_xZr_{1-x}B_{12}$. The integration was carried out in the temperature range from $T$ to $T_c$. Before integration the specific heat data in the normal and superconducting states were approximated by polynomials of the 4$^{th}$ order. The $H_{cm}(0)$ values were obtained by extrapolation of $H_{cm}(T)$ curves in the framework of the standard Bardeen-Cooper-Schriffer (BCS) relation

$$H_{cm}(T)/H_{cm}(0) = 1.7367(1-T/T_c)[1-0.327(1-T/T_c)-0.0949(1-T/T_c)^2] \qquad (3.4)$$

and $H_{c2}(0)$ magnitudes were defined within the framework of the phenomenological polynomial approximation.

**Section 4. Superconductive state characteristics.**

For independent evaluation of the coherence length $\xi(0)$ and Ginzburg-Landau-Maki parameter $\kappa_1(T)$ we use standart BCS relations

$$\xi(0)=(\Phi_0/2\pi H_{c2}(0))^{1/2} \qquad (4.1)$$

$$\kappa_1(T)=2^{-1/2} H_{c2}(T)/H_{cm}(T) \qquad (4.2)$$

where $\Phi_0$ denotes the flux quantum.

Two-band $\alpha$-model was applied to describe the electronic specific heat in the superconducting state:

$$C_{si}(T)=A_0 T^{-3/2} exp(-\Delta_i(0)/k_B T) \qquad (4.3)$$

($i = 1, 2$; $A_0$ is the temperature independent coefficient; $\Delta_i(0)$ is superconducting gap in the $i$-band, and $k_B$ is Boltzmann constant).